\documentclass[11pt, journal,onecolumn]{IEEEtran}
\IEEEoverridecommandlockouts
\usepackage{cite}
\usepackage{amsmath,amssymb,amsfonts}
\usepackage{booktabs}
\usepackage{graphicx}
\usepackage{textcomp}
\usepackage{xcolor}
\usepackage{graphicx}
\usepackage{amsmath, xspace,  comment}
\usepackage{tikz, xcolor}
\usepackage{algpseudocode}
\usepackage[linesnumbered]{algorithm2e}
\usepackage{colortbl}
\usepackage{textcase} 
\usepackage{enumitem}
\usepackage{balance}
\usepackage{graphicx}
\usepackage{multirow}
\usepackage{hhline}
\usepackage{caption}
\usepackage[caption=false]{subfig}
\usepackage{soul}
\usepackage{hyperref}
\usepackage{makecell}
\usepackage[a4paper,left=1.2in,top=1.5in,right=1in,bottom=1in,nohead]{geometry}

\def\BibTeX{{\rm B\kern-.05em{\sc i\kern-.025em b}\kern-.08em
    T\kern-.1667em\lower.7ex\hbox{E}\kern-.125emX}}

\let\svthefootnote\thefootnote
\newcommand\freefootnote[1]{%
\let\thefootnote\relax%
\footnotetext{#1}%
\let\thefootnote\svthefootnote%
}
\newcommand*\cib[1]{\tikz[baseline=(char.base)]{
                            \node[shape=circle,fill=black,text=white,draw,inner sep=0.3pt] (char) {#1};}}

\newcommand{\etal}{{\em et al.}\xspace}
\newcommand{\eg}{{\em e.g.,}\xspace}
\newcommand{\ie}{{\em i.e.,}\xspace}

\newcommand{\BfPara}[1]{{\vspace{1ex}\noindent{\bf \em#1.}}\xspace}
\usepackage{tcolorbox}
\tcbuselibrary{theorems}

\newtcbtheorem[]{observation}{\textbf{Observation}}%
{colback=white,colframe=white!10!black,fonttitle=\bfseries,coltitle=white}{}

\hypersetup{
	colorlinks=true,
	urlcolor=blue!70!black,
	linkcolor=red!70!black,
	citecolor=purple,
}

\newcommand{\UlPara}[1]{{\noindent\bf{\ul{#1.}}}\xspace}

\setul{0.8pt}{}
\begin{document}

\title{From Attack to Defense: Insights into \\Deep Learning Security Measures in Black-Box Settings}

\author{
\IEEEauthorblockN{Firuz Juraev$^{1}$, Mohammed Abuhamad$^{2}$, Eric Chan-Tin$^{2}$, \\George K. Thiruvathukal$^{2}$, and Tamer Abuhmed$^{1}$\\}
\IEEEauthorblockA{$^{1}$Department of Computer Science and Engineering, Sungkyunkwan University\\
$^{2}$ Department of Computer Science, Loyola University Chicago\\
Email: fjuraev@g.skku.edu, mabuhamad@luc.edu, chantin@cs.luc.edu,\\ gkt@cs.luc.edu, tamer@skku.edu  
}}

\maketitle 

\begin{abstract}

Deep Learning (DL) is rapidly maturing to the point that it can be used in safety- and security-crucial applications, such as self-driving vehicles, surveillance, drones, and robots. 
However, adversarial samples, which are undetectable to the human eye, pose a serious threat that can cause the model to misbehave and compromise the performance of such applications. 
Addressing the robustness of DL models has become 
crucial to understanding and defending against adversarial attacks.
In this study, we perform comprehensive experiments to examine the effect of adversarial attacks and defenses on various model architectures across well-known datasets. 
Our research focuses on black-box attacks such as \textit{SimBA}, \textit{HopSkipJump}, \textit{MGAAttack}, and \textit{boundary attacks}, as well as preprocessor-based defensive mechanisms, including \textit{bits squeezing}, \textit{median smoothing}, and \textit{JPEG filter}. 
Experimenting with various models, our results demonstrate that the level of noise needed for the attack increases as the number of layers increases.
Moreover, the attack success rate decreases as the number of layers increases. 
This indicates that model complexity and robustness have a significant relationship. 
Investigating the diversity and robustness relationship, our experiments with diverse models show that having a large number of parameters does not imply higher robustness.

Our experiments extend to show the effects of the training dataset on model robustness. 
Using various datasets such as ImageNet-1000, CIFAR-100, and CIFAR-10 are used to evaluate the black-box attacks. Considering the multiple dimensions of our analysis, \eg model complexity and training dataset, we examined the behavior of black-box attacks when models apply defenses. Our results show that applying defense strategies can significantly reduce attack effectiveness. This research provides in-depth analysis and insight into the robustness of DL models against various attacks, and defenses\freefootnote{The code, models, and detailed settings are available at \hyperlink{https://github.com/InfoLab-SKKU/Adversarial-Attacks-Analysis}{\textit{\textcolor{purple}{https://github.com/InfoLab-SKKU/Adversarial-Attacks-Analysis}}}
\\Corresponding author: Tamer Abuhmed (tamer@skku.edu)}.

\end{abstract}
\thispagestyle{plain}
\section{Introduction}
Deep Learning (DL) has grown into a powerful and effective tool that contributes to solving a wide range of complicated learning tasks that were previously impossible to solve using conventional Machine Learning (ML) approaches. 
DL has been extensively used in a wide variety of modern day-to-day applications \cite{abuhamad2021large,ali2022effective}, and it has made tremendous progress in terms of equal or beyond human-level performance. 

Consequently, DL-based solutions are widely used in safety-, privacy-, and security-critical situations, such as malware detection \cite{singh2021classification}, self-driving vehicles, drones, personal assistants, smart homes, and robots. 

Despite the success of DL, concerns have been raised regarding efficiency, energy consumption, data and model privacy, and susceptibility to adversarial examples. %
After the findings of Szegedy \etal \cite{szegedy2014intriguing}, several intriguing studies appeared in the research community involving adversarial attacks on DL. 
For instance, vulnerabilities of automatic speech recognition \cite{carlini2016hidden}, voice controllable systems \cite{zhang2017dolphinattack}, and autonomous vehicles \cite{kurakin2016adversarial} have been demonstrated by adversarial attacks. 

Existing adversarial attacks can be classified into three categories: \textbf{white-}, \textbf{gray-}, and \textbf{black-box attacks}. 
The adversaries' knowledge is what distinguishes the three attacks. 
Generally, DL models are of black-box nature, and black-box attacks are more realistic in many cases and are closer to real-world challenges. 

Black-box attacks target the model with no knowledge of the systems' inner processes. Such attacks are relatively difficult to conduct and alarming-to-devastating when they succeed \cite{10397075,abdukhamidov2023unveiling}. 
We shed light on the characteristics, effects across multiple dimensions, and defenses against black-box attacks. In this work, we focus on the most frequently accessible methods among black-box attacks, such as Simple Black-box Attack (\textbf{SimBA}) \cite{guo2019simple}, \textbf{HopSkipJump} \cite{chen2020hopskipjumpattack}, Microbial Genetic Algorithm-based Attack (\textbf{MGAAttack}) \cite{wang2020mgaattack} and \textbf{boundary attacks} \cite{brendel2018decision}. We study the behavior of these attacks in different settings and various models.

Recent studies of defensive strategies against adversarial attacks suggest a plethora of trainer-, detector-, and preprocessor-type defenses. 
Unlike trainer- and detector-type defenses, preprocessor-type defenses neither optimize the model's weight nor change the model's architecture. Instead, the adversarial input is preprocessed to make it less-to-non-adversary. This work also investigates the effects of commonly used preprocessor-type defenses (\eg bit squeezing \cite{xu2018feature}, median smoothing \cite{osadchy2017no}, and JPEG filter \cite{dziugaite2016study}) and Adversarial training \cite{shafahi2019adversarial} on black-box attacks.

In addition, this work examines the \ul{complexity}, \ul{diversity}, and \ul{robustness} relationship among various DL model families. In the case of \ul{complexity-robustness} relationship, we study \textbf{ResNet}, \textbf{VGG}, and \textbf{DenseNet} families. For the \ul{diversity-robustness} relationship, we explore \textbf{InceptionV3}, \textbf{Xception}, \textbf{GoogleNet}, \textbf{MobileNetV2}, and \textbf{ShuffleNetV2}. 
Interestingly, it has long been assumed that larger models with more trainable parameters outperform smaller models because they can capture more complex patterns in the data.
However, our in-depth analysis and experiments on the relationship between the complexity and robustness of models show that the number of parameters in a model is not a concrete indicator of robustness. In fact, our results indicate that middle-size models may be more robust against adversarial attacks compared to heavy-weight models.
For example, models like VGG-19 and ResNet-152, which have a large number of parameters (143,667,240 and 60,192,808, respectively), can be more vulnerable to attacks compared to middle-weight models such as Xception (22,855,952 parameters) and GoogLeNet (6,624,904 parameters).
On the other hand, our results suggest that light-weighted models like MobileNet V2 and ShuffleNet V2 (3,504,872 and 2,278,604 parameters, respectively) can also be less robust compared to larger models.
This study highlights the importance of taking steps to ensure that the model (of any size) has sufﬁcient generalization and robustness capabilities.

This research also investigates the effects of using different training datasets on attacking various models. 
Although the sensitivity of models to adversarial examples in the image context and throughout the literature was typically assessed using specific datasets,
Carlini \etal \cite{carlini2017adversarial} showed that DL approaches developed using some datasets may not always extend to other datasets. For example, CIFAR {\cite{krizhevsky2009learning}} and MNIST {\cite{lecun1998gradient}} datasets, compared to ImageNet \cite{russakovsky2015imagenet}, include images with lower resolution and fewer categories. 
Therefore, most of the research efforts recently have favored ImageNet over CIFAR and MNIST \cite{guo2019simple}. 
Our analysis is done on three datasets, \ie ImageNet-1000, CIFAR-100, and CIFAR-10.

\BfPara{Contributions} In this work, we investigate the various aspects of black-box attacks and preprocessor-type defenses on the state-of-the-art models, including \textbf{ResNet}, \textbf{VGG}, \textbf{DenseNet} families using benchmark datasets such as ImageNet-1000, CIFAR-100, and CIFAR-10. 
Furthermore, different architectures are examined for the diversity-robustness relationship of models including, \textbf{InceptionV3}, \textbf{Xeption}, \textbf{GoogleNet}, \textbf{MobileNetV2}, and \textbf{ShuffleNetV2}. In order to address the following research questions, large-scale experiments are performed and analyzed:  
\begin{enumerate}[label={},itemindent=-2.8ex, leftmargin=*]
    \item \cib{1} The influence of model complexity on the effectiveness of various black-box attacks remains unclear. Specifically, {\em Does the model's complexity impact the attack's success/failure rate?} To shed light on this aspect, our study investigates the relationship between the complexity and robustness of models (from multiple DL families) across various black-box attacks and datasets.
    \item \cib{2} We investigate whether there exists a relationship between the adopted architecture/design and the robustness of models against adversarial attacks. Specifically, \emph{do model design choices impact robustness?}
    We examine the behavior and vulnerability of diverse models under adversarial conditions.
    \item \cib{3} 
    {\em When adversarial examples are provided to the model, how do models address the attack problem while implementing defenses?} The most commonly used preprocessor defense strategies are analyzed to address this question.
\end{enumerate}

\BfPara{Organization} The remainder of the paper is organized as follows: 

Section \ref{sec:methods} highlights the methods and the essential principles for black-box attacks and defensive strategies. 

Section \ref{sec:results} discusses the experimental findings and analysis. The results are summarized in Section \ref{sec:discussion} and the open challenges in attacks and defenses. The paper is concluded in Section \ref{sec:conclusion}.

\section{Experimental Settings} \label{sec:methods}
This section highlights the notation used in this study, describes the attacks and defenses, and outlines the settings of the experiment.

\subsection{Black-box Attack}

This research examines the most predominant black-box attacks, which are summarized as follows.

\BfPara{SimBA \cite{guo2019simple}} This attack aims to detect a modest disturbance $\sigma$ that causes the prediction $\mathcal{F}(x+\sigma)$ to differ from the actual value $y$. It may be defined as a modified restricted optimization problem \cite{guo2019simple}, with a small misapplication of the notation:

\begin{equation*}
    \begin{split}
        \min_{(x+\sigma)} & \hspace{0.2cm} \mathcal{M}\big[(x+\sigma),~~ x \big]\\
        \text{subject to} \hspace{0.1cm} & \hspace{0.2cm}  ||\sigma||_2 < \omega, \quad \text{queries} \leq \mathcal{B}, \\
     \end{split}
    \label{eq:4}
\end{equation*}

\noindent where the distance metric $\mathcal{M}$ is minimized using a perturbation $\sigma$ with $\ell_2$ below a pre-defined threshold $\omega$ when the adversary is allowed to query the model less than $\mathcal{B}$ times.

\vspace{1.5ex}
\BfPara{HopSkipJump \cite{chen2020hopskipjumpattack}} This attack generates adversarial examples by looking at the output labels $y'$ of the targeted model. The attack is divided into two modes: Targeted (\textbf{T}) and Untargeted (\textbf{UT}) attacks \cite{chen2020hopskipjumpattack}. The objectives of both settings are as follows.

\begin{equation*}
    \begin{split}
         & \min_{x^*} \mathcal{M}(x^*,x) \\
        \textbf{T:} & \hspace{0.2cm}   \text{subject to} \hspace{0.1cm} \mathcal{F}(x^*)=y' \\
        \textbf{UT:} & \hspace{0.2cm}  \text{subject to} \hspace{0.1cm} \mathcal{F}(x^*) \neq \mathcal{F}(x) \neq y \\
    \end{split}
    \label{eq:5}
\end{equation*}

\noindent where $x^* = (x+\sigma)$ and $y'$ is the target class. The attack suggests the following boundary function $\mathcal{S}(x^*)$:
\begin{equation*}
    \begin{split}
        \textbf{T:} & \hspace{0.2cm} \hspace{0.1cm} \mathcal{F}(x^*)_y - \max_{y' \neq y} \mathcal{F}(x^*)_{y'} \\
        \textbf{UT:} & \hspace{0.2cm} \hspace{0.1cm} \max_{y' \neq \mathcal{F}(x)} \mathcal{F}(x^*)_{y'} - \mathcal{F}(x^*)_{\mathcal{F}(X)} \\
    \end{split}
    \label{eq:6}
\end{equation*}

\BfPara{MGAAttack \cite{wang2020mgaattack}} This method involves the five steps of genetic algorithms: initialization, selection, crossover, mutation, and population update. Transfer-based attacks are used to initialize the population $\delta_i$ and its quality is evaluated by a fitness function. The population passes on their genetic information with high fitness by crossover. The mutation step is crucial to maintaining diversity in the population of adversarial perturbations and to avoiding getting stuck in local optima. To achieve this, the mutation operator in MGAAttack randomly modifies the existing perturbations in the population. Specifically, the mutation operator applies a masking operation (denoted as $\mathsf{MASK}_{mr}$) to a portion of the existing perturbation (denoted as $\delta_i$) and then updates it by combining it with the original perturbation. The masking operation randomly selects a subset of the perturbation and sets its values to zero, which creates a mask of the same shape as the perturbation. The mask is then used to modify the perturbation by setting the masked values to their negative values and the non-masked values to their original values, as shown in the equation:
\begin{equation*}
    \delta_i = -\delta_i * \mathsf{MASK}_{mr} + \delta_i * (1-\mathsf{MASK}_{mr}),
\end{equation*}

{This operation flips the sign of the selected values, effectively changing the direction of the perturbation. By applying this operation with a certain probability during the mutation step, MGAAttack can explore new regions of the search space and avoid being trapped in local optima.}

\BfPara{Boundary Attack \cite{brendel2018decision}} This is a type of adversarial attack method that is designed to find the smallest possible perturbation to an input image that will cause the model to misclassify it. This method works by constraining the input image to a range of [0, 255] and then resampling it to a uniform distribution $U$[0, 1].

The effectiveness of the attack is determined by the resampled distribution, which is used to generate the perturbations. At each stage of the attack, a perturbation $\delta^k$ is selected from a maximum entropy distribution designed to maximize the uncertainty of the model prediction. The perturbation is then added to the original image to generate a perturbed image.

The goal of the attack is to decrease the distance between the perturbed image and the original image by a factor of $\eta$, where $\eta$ is a hyperparameter that controls the magnitude of the perturbation. This is achieved by iteratively selecting perturbations that move the perturbed image closer to the decision boundary of the model. 

\begin{equation*}
        \mathcal{F}(x,x^*) - \mathcal{F}(x,x^*+\delta^k) = \hspace{0.1cm} \eta . \mathcal{F}(x,x^*),
    \label{eq:7}
\end{equation*}

The equation represents the difference in the prediction of the model between the original image $x$ and the perturbed image $x$*+$\delta^k$. The left-hand side of the equation measures the change in the prediction of the model due to the addition of the perturbation $\delta^k$. The right-hand side of the equation represents the magnitude of the perturbation $\delta^k$, scaled by the hyperparameter $\eta$.

By minimizing this equation, the attack algorithm can find the smallest possible perturbation that will cause the model to misclassify the image. The algorithm repeats this process until the model's prediction changes, indicating that a successful adversarial example has been found.

\subsection{Defense Mechanisms}

In addition to our analysis of adversarial attacks, we investigated the impact of employing multiple preprocessor-type defense strategies and adversarial training. 
Using preprocessor-type defenses in black-box settings offers several advantages. For example, preprocessor-type defenses can be applied to any deep learning model without requiring changes to the model architecture, training process, or/and the use of auxiliary modules. This makes preprocessor-type defenses more practical and scalable compared to other defense techniques.
Moreover, preprocessor-type defenses can often be applied efficiently, without significant computational overhead.
Studying the behavior of models while applying preprocessor-type defenses against various black-box attacks can provide insight into the strengths and limitations of such defenses. We also show how adversarial training impacts the behavior of attacks.
The defense strategies utilized in this study are summarized as follows.

\BfPara{Bits Squeezing \cite{xu2018feature}} This defense removes redundant pixels in the color bit depths. 
The bit squeezing defense technique reduces the input image to the i-bit depth $(1\leq i\leq7)$ by first multiplying the input with \textit{round}$(2^i-1)$. The output is normalized to $[0,1]$, by dividing $2^i-1$. The information capacity of the representation is reduced from 24-bit to i-bit with the integer-rounding operation.

\BfPara{Median Smoothing \cite{osadchy2017no}} This defense is particularly successful against adversarial examples, since it reduces noise by smoothing pixel values using neighboring pixels. A sliding window is applied to each pixel, replacing the center pixel with the median value of the nearby pixels inside the window. 

\BfPara{JPEG Filtering \cite{dziugaite2016study}} This defense is used to reverse small adversarial perturbations induced by black-box attacks. Let the weights of a DNN be $\Theta$, the noise is calculated as follows:
\begin{equation*}
\Gamma(x) = \frac{\lambda}{255} \hspace{0.2cm} \textbf{JPEG}\Big(\nabla_{x^*} \mathcal{J}(x^*,\Theta, y) \Big|_{x^*=x, y=l_x} \Big), 
\end{equation*}

\noindent where $\textbf{JPEG}(.)$ is the filtering process, and $l_x= \text{argmax}~ \mathcal{P}_\Theta(y|x)$ that represents the class. The gradient $\nabla_{x^*} \mathcal{J}(.)$ can be calculated with backpropagation.

\BfPara{Adversarial training \cite{shafahi2019adversarial}} This is a technique used in machine learning to improve the robustness of a model against adversarial attacks. It involves training the model with adversarial examples generated by an attacker to increase the model's ability to detect and resist such attacks. This approach has been successful in improving the accuracy and security of machine learning models.

\begin{table*}[h]
 \centering
\caption{{{The accuracy of models using} {different datasets for \ul{Exp 1}.}}}
\label{tab:model_acc_1}
\resizebox{0.65\textwidth}{!}{%
\begin{tabular}{l|ccc}
\toprule
\textbf{Model}  & \makecell{\textbf{Top 5\% Accuracy}\\ \textbf{(ImageNet)}} & \makecell{\textbf{Accuracy}\\ \textbf{(CIFAR 100)}} & \makecell{\textbf{Accuracy}\\ \textbf{(CIFAR 10)}} \\  
\midrule
ResNet 18   & 89.08\%   & 75.94\%    & 94.79\%       \\
ResNet 34   & 91.42\%   & 76.72\%   & 94.77\%       \\
ResNet 50   & 92.86\%   & 76.96\%   & 95.30\%       \\
ResNet 101  & 93.55\%   & 76.22\%   & 94.99\%       \\
ResNet 152  & 94.05\%   & 74.88\%   & 94.60\%       \\ \midrule
VGG 11   & 88.63\%   & 71.08\%   & 91.94\%       \\
VGG 13   & 89.25\%   & 73.50\%   & 93.70\%       \\
VGG 16   & 90.38\%   & 72.86\%   & 93.30\%       \\
VGG 19   & 90.88\%   & 71.84\%   & 93.18\%       \\ \midrule
DenseNet 121   & 91.97\%   & 77.54\%   & 95.06\%      \\
DenseNet 169   & 92.81\%   & 78.67\%   & 95.10\%      \\
DenseNet 201   & 93.37\%   & 78.63\%   & 95.17\%      \\ 
\bottomrule
\end{tabular}%
}
\end{table*}

\begin{table*}[h]
\centering
\caption{{{The accuracy of models using} {different datasets for \ul{Exp 2}.}}}
\label{tab:model_acc_2}
\resizebox{0.65\textwidth}{!}{%
\begin{tabular}{l|ccc}
\toprule
\textbf{Model}  & \makecell{\textbf{Top 5\% Accuracy}\\ \textbf{(ImageNet)}} & \makecell{\textbf{Accuracy}\\ \textbf{(CIFAR 100)}} & \makecell{\textbf{Accuracy}\\ \textbf{(CIFAR 10)}} \\  
\midrule
VGG 19   & 90.88\%   & 71.84\%   & 93.18\%       \\      
ResNet 152  & 94.05\%   & 74.88\%   & 94.60\%       \\
DenseNet 169   & 93.56\%   & 79.05\%   & 95.43\%      \\ \midrule
Inception V3   & 93.45\%   & 77.12\%   & 94.77\%      \\
Xception   & 94.29\%   & 74.67\%   & 93.49\%      \\ 
GoogLeNet   & 89.53\%   & 75.93\%   & 95.01\%      \\ \midrule
MobileNet V2	 & 90.28\%   & 69.44\%   & 94.20\%      \\ 
ShuffleNet V2   & 	88.31\%   & 64.19\%   & 90.04\%      \\ \bottomrule
\end{tabular}%
}
\end{table*}

\begin{table*}[h]
\caption{{{The parameters and number } {of layers of selected models for \ul{Exp 2}.}} }
\label{tab:model_info}
\centering
\resizebox{0.65\textwidth}{!}{%
\begin{tabular}{l|ccc}
\toprule
\textbf{Model}  & \makecell{\textbf{Number of}\\ \textbf{parameters}} & \makecell{\textbf{Number of}\\ \textbf{Layers}} & \makecell{\textbf{Activation}\\ \textbf{functions}} \\  
\midrule
VGG 19               & 143,667,240                   & 19                        & ReLU                          \\
ResNet 152           & 60,192,808                    & 152                       & ReLU                          \\
DenseNet 169         & 14,149,480                    & 169                        & ReLU                          \\ \midrule
Inception V3         & 27,161,264                    & 48                        & ReLU                          \\
Xception             & 22,855,952                    & 71                        & ReLU                          \\
GoogLeNet            & 6,624,904                     & 22                        & ReLU                          \\ \midrule
MobileNet V2         & 3,504,872                     & 53                        & ReLU                          \\
ShuffleNet V2        & 2,278,604                     & 65                        & ReLU                     \\ \bottomrule
\end{tabular}%
}
\end{table*}

\begin{figure*}[t]
    \centering
    \includegraphics[width=0.97\linewidth]{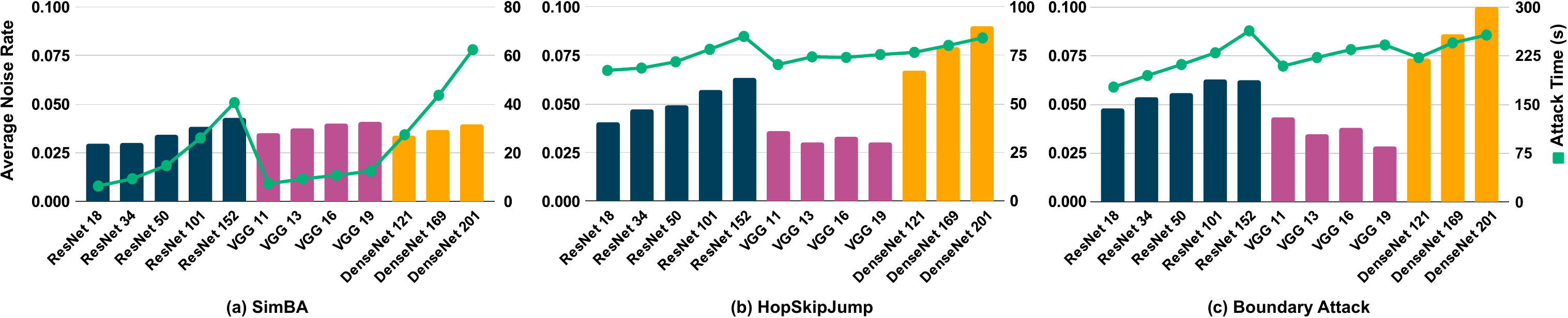} 
    \caption{Relation between the number of layers and the amount of noise and time needed (green line) for an attack to succeed. Noise rate and time increase as the number of layers increases in models of different families in three black-box attacks.
    }
    \label{fig:experiment1_diagram} 
    
\end{figure*}

\BfPara{Dataset} ImageNet \cite{russakovsky2015imagenet}, CIFAR-10 \cite{krizhevsky2009learning}, and CIFAR-100 \cite{krizhevsky2009learning} are three common datasets used in all experiments. The ImageNet dataset, with a $224 \times 224$ image size and 14 million samples, is organized into 1000 classes. 
The CIFAR-100 dataset, with a $32 \times 32$ image size and 60,000 samples, is divided into 100 classes. 
The CIFAR-10 dataset contains 60,000 images with a $32 \times 32$ image size for 10 classes.
Due to the time complexity of the black-box attacks, all of the results are centered on test data of 1000 images. These images are selected based on two conditions: \cib{1} one example from each class of ImageNet, and \cib{2} each image must be classified correctly by all selected models. In both CIFAR-100 and CIFAR-10 datasets, a balanced sampling approach was used to select an equal number of images from each respective class, and all selected images are classified correctly by all selected models.

\BfPara{Models} To examine the complexity and robustness relationship (Section \ref{sec:ComplexityRobustness}), 12 models from the ResNet \cite{he2016deep}, VGG \cite{simonyan2014very}, and DenseNet families are employed. In the second set of experiments, seven distinct models are trained on each dataset, including GoogLeNet, InceptionV3, Xception, MobileNetV2, and ShuffleNetV2, to explore the model diversity-robustness relationship (Section \ref{sec:DiversityRobustness}). 
All of these models are used across our sets of experiments to explore the behavior of attacks in different settings.
Pre-trained models for the ImageNet-1000 dataset based on the Pytorch framework were employed. {We trained all models for the CIFAR-100 and CIFAR-10 datasets from scratch.} {For both CIFAR-100 and CIFAR-10 datasets, the training continues for 200 epochs with a batch size of 128 and a learning rate of 0.01. The accuracy of all models across datasets is shown in Tables \ref{tab:model_acc_1} and \ref{tab:model_acc_2}.}

\BfPara{Evaluation Metrics} The used metrics are: \cib{1} \textit{Misclassification confidence (MC)} \cite{zhang2020interpretable}: the probability that a model assigns incorrectly to classify an adversarial image. \cib{2} \textit{Attack time}: the estimated time taken to carry out a successful attack. We presented an average attack time in seconds for each model.  \cib{3} \textit{Noise rate}: the degree of noise is determined using the Structural Similarity Index Measure (SSIM) \cite{wang2004ssim}. SSIM measures how two images are similar to each other. To calculate the noise rate, we subtracted the SSIM value from 1 (\ie $noise \hspace{0.1cm} rate = 1 - SSIM$). 

\cib{4} \textit{Attack success rate} \cite{zhang2020interpretable}: the ratio of successful trials to total trials. \cib{5} \textit{Confidence score}: the model probability of assigning an input to an output.

\BfPara{Experiment Workstation} 
 The experiments are conducted on a machine equipped with an Intel Xeon(R) CPU E5-2620 v3 @ 2.40 GHz$\times$ 24 with Cuda-10.0 and three GEFORCE RTX 2080 Ti 12 GB GPUs, as well as Python 3.7.7 distributed in Anaconda 4.8.3 (64-bit).

\BfPara{Attack algorithms settings}
The used adversarial attacks: \cib{1} \textit{Simple Black-box Attack (SimBA)} with the following parameters: epsilon is 0.5 (overshoot parameter), maximum iterations are 6000. \cib{2} \textit{HopSkipJump Attack} with the following parameters: batch size is 64, and maximum iterations are 20. \cib{3} \textit{Boundary Attack} with the following parameters: epsilon is 0.01, and maximum iterations are 1000. \cib{4} \textit{MGA Attack} with the following parameters: epsilon is 0.047, maximum queries are 1000, and mutation rate is 0.001. The parameters of attacks are tuned to get the maximum attack success rate.

\BfPara{Defense algorithms settings} 
The used defense algorithms: \cib{1} \textit{Bits Squeezing} with the parameter of bit depth = 4. \cib{2} \textit{Median Smoothing} with the parameter of kernel size as 2. \cib{3} \textit{JPEG Filter} with the parameter of quality as 75. The parameters of defense algorithms are selected with minimal settings to keep the image quality.

\section{Results and Analysis} \label{sec:results}
Adversarial perturbations are intriguing because
the adversaries are quite observable to DNNs over a broad range of training regimens. First, we must distinguish between the model's complexity and diversity. The number of layers employed in the network design is used to describe \textit{model's complexity}. On the other hand, \textit{model diversity} is characterized as distinct network architectures.

In this section, the findings of the experiments are discussed.

\subsection{{Exp 1}--- Model Complexity and Robustness} \label{sec:ComplexityRobustness}
The ImageNet dataset is used in this set of experiments to explore the complexity-robustness relationship of models. The complexity is implied by the depth and breadth of models, \ie the number of layers and the number of parameters. The robustness of models is measured by their ability to maintain high accuracy on the test set when attacked by adversarial examples. These experiments aim to explore whether there is a trade-off between model complexity and robustness.
For a fair comparison, the same family of models is utilized with a varying number of layers, \eg ResNet family (ResNet-18, ResNet-34, ResNet-50, ResNet-101, and ResNet-152), VGG (VGG-11, VGG-13, VGG-16, and VGG-19), and DenseNet (DenseNet-121, DenseNet-169, and DenseNet-201). The experiments are repeated 10 times on the same selected image set to obtain more stable results.

\begin{figure}[t]
    \centering
    \includegraphics[width=0.99\linewidth]{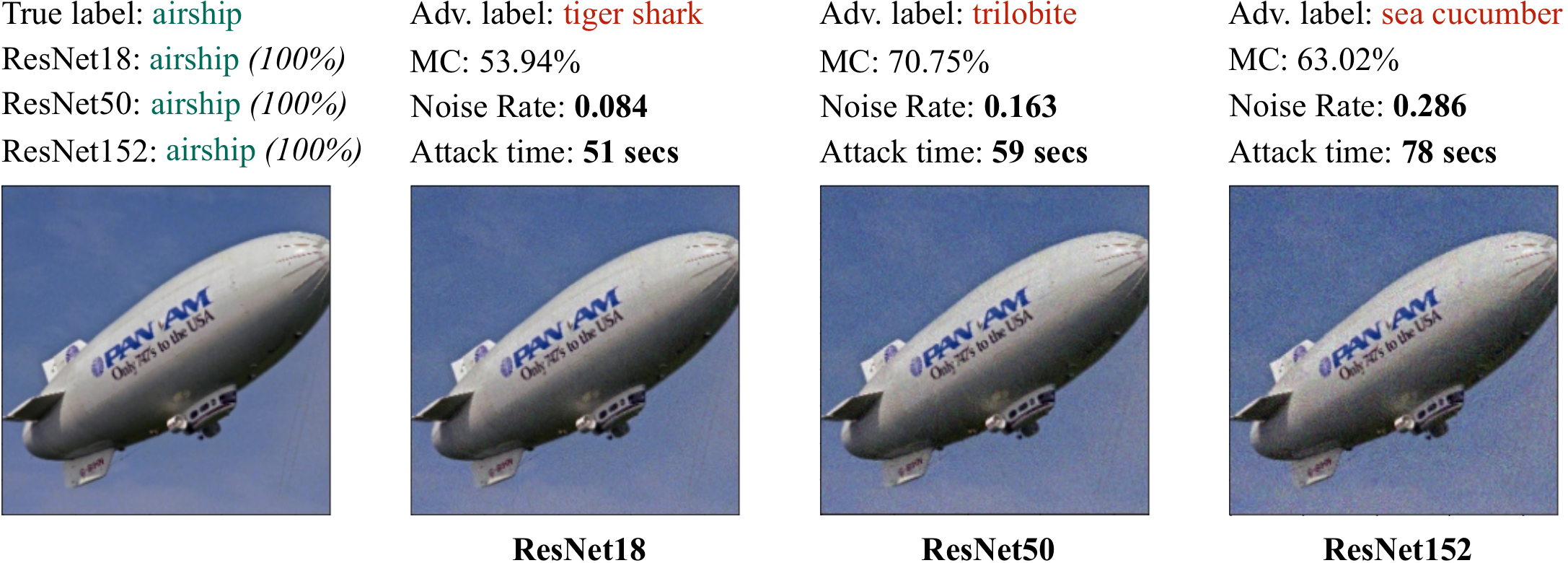} 
    \caption{Noise, Misclassification confidence (MC), and time needed by the HopSkipJump attack for ResNet-18, ResNet-50, and ResNet-152 models.}
    \label{fig:adv_example} 
\end{figure}

\UlPara{Complexity via Model's Depth} Figure  \ref{fig:experiment1_diagram} shows that as the number of layers increases, the noise rate steadily increases, indicating that a model becomes more resistant to adversarial attacks as the number of layers increases. Furthermore, successfully attacking deeper models requires more time and computation. The figure also shows that the time (green line) needed to generate adversarial samples increases for deeper models within a family.
The varying noise and attack times are evident across models from ResNet and DenseNet with a varying number of layers. However, because there is less variation in the number of layers of VGG models, the variance is less obvious.
This suggests that as models become more complex (depth-wise), generating effective adversarial attacks becomes increasingly difficult and resource-intensive. 
For example, Figure \ref{fig:adv_example} the correct label for a benign image is an airship. Attacking ResNet-18 requires 8.4\% adversarial noise using the HopSkipJump attack in 51 seconds, while attacking ResNet-50 and ResNet-152 requires 16.3\% and 28.6\% adversarial noise in 59 and 78 seconds, respectively.

\UlPara{Complexity via Model's Parameters} On the other hand, the complexity of the models in terms of the number of parameters might suggest different results. For example, using HopSkipJump and Boundary Attack (Figure \ref{fig:experiment1_diagram}-(b) and -(c)), VGG models (\eg VGG-19 with 143,667,240 parameters) are more susceptible to perturbations than ResNet (\eg ResNet-152 with 60,192,808 parameters) and DenseNet (\eg DenseNet-169 with 14,149,480 parameters). This also holds true when comparing results from the ResNet and DenseNet families, as DenseNet models are more resilient to perturbations.  

These results indicate that the model complexity (\ie number of layers and parameters)---alone does not necessarily determine the susceptibility to adversarial attacks and that the architectural design itself impacts the robustness. This encourages the following set of experiments to explore other factors, \eg model design.

\begin{figure}[t]
    \centering
    \includegraphics[width=.65\linewidth]{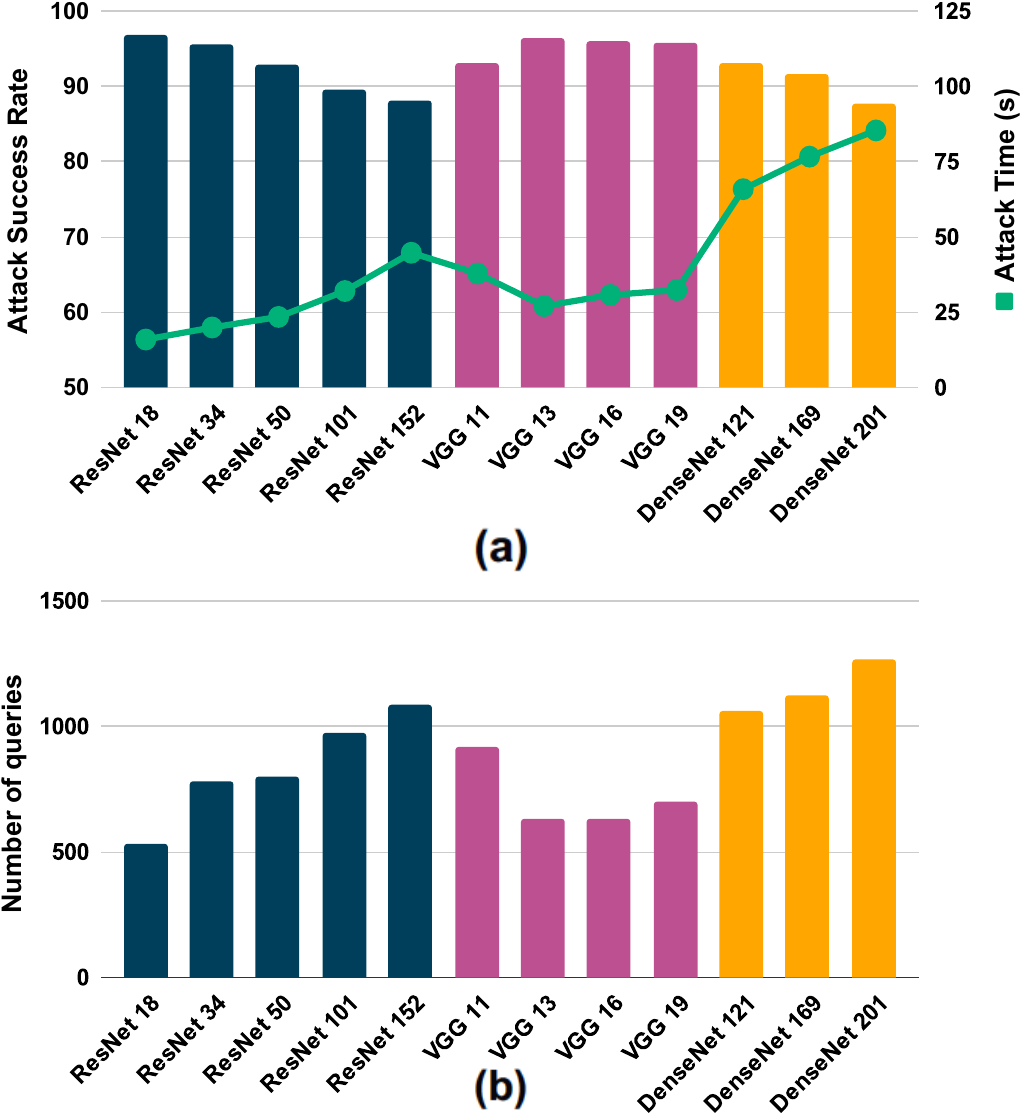} 
    \caption{The amount of noise and time needed (green line) for MGAAttack to succeed against various models. As the number of layers increases, the attack success rate decreases (above), and the number of queries increases (below).}
    \label{fig:experiment1_mga_attack} 
  
\end{figure}

\UlPara{Complexity against Sophisticated Attacks}
MGAAttack adopts a distinct approach from previous attacks that use heuristics to generate adversarial samples with minimal noise through an evolutionary approach.  This approach, among many emerging sophisticated methods, allows a more efficient and effective way of creating adversarial samples. Therefore, we investigated the relationship between model complexity and robustness against MGAAttack through two key metrics, namely attack success rate and attack time. 
Figure \ref{fig:experiment1_mga_attack} illustrates that as the number of layers in the models increases, the success rate of the attack decreases and the number of queries for the attack increases. In terms of the number of parameters, similar observations as from previous experiments were made, where models with a larger number of parameters tended to be more susceptible to attacks.

  \begin{observation*}{Number of layers affects robustness}{}
  The model becomes more robust as the number of layers on the same model increases (See Figure \ref{fig:experiment1_diagram}). The black-box attacks cause more noise and time as the model (\ie from the same family) grows more intricate.
\end{observation*}

\begin{figure}[t]
    \centering
    \includegraphics[width=0.65\linewidth]{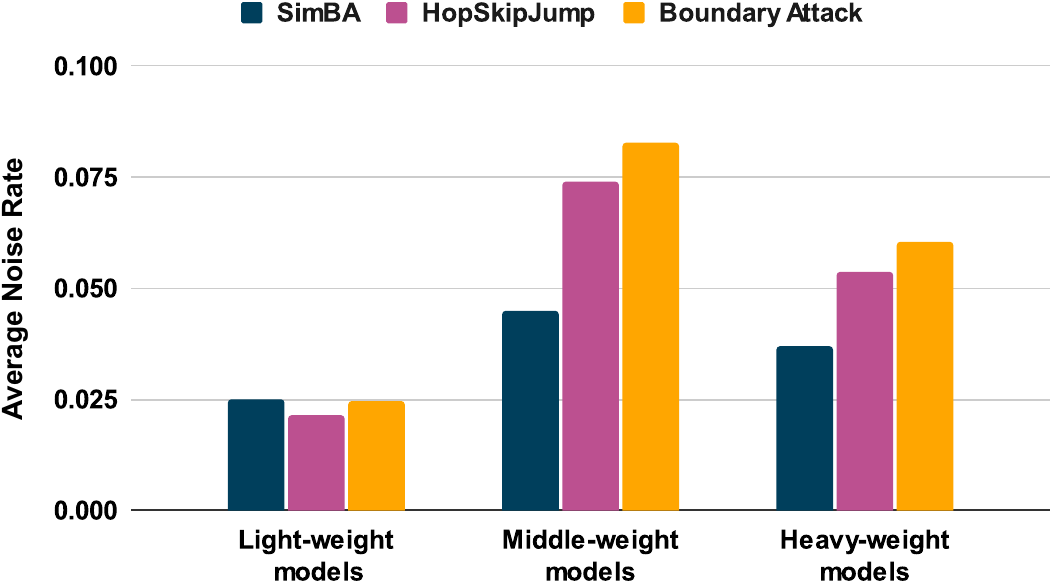}
    \caption{Attack needed noise across light-weight, middle-weight, and heavy-weight models when using SimBA, HopSkipJump, and Boundary attacks.}
    \label{fig:experiment2_diagram} 
\end{figure}

\begin{table}[]
\caption{Attack added noise across 8 diverse models when using SimBA, HopSkipJump, and Boundary attacks.}
\label{tab:experiment2_detail}
\centering
\resizebox{0.8\textwidth}{!}{%
\begin{tabular}{l|l|c|c|c}
\rowcolor[HTML]{CCCCCC} 
\textbf{Model Type}                      & \textbf{Target Model}  & \textbf{SimBA}       & \textbf{HopSkipJump} & \textbf{Boundary Attack} \\ \hline
& VGG 19      & 0.035±0.041 & 0.024±0.036 & 0.027±0.039     \\
& ResNet 152  & 0.039±0.043 & 0.055±0.061 & 0.062±0.069     \\
\multirow{-3}{*}{\textbf{Heavy-weight}}  & DenseNet 161  & 0.036±0.040 & 0.081±0.077 & 0.091±0.086     \\ \hline
& Inception V3  & 0.038±0.046 & 0.058±0.071 & 0.071±0.091     \\
& Xception      & 0.050±0.048 & 0.083±0.080 & 0.088±0.091     \\
\multirow{-3}{*}{\textbf{Middle-weight}} & GoogLeNet     & 0.045±0.044 & 0.081±0.088 & 0.088±0.093     \\ \hline
& MobileNet V2  & 0.029±0.036 & 0.028±0.038 & 0.031±0.044     \\
\multirow{-2}{*}{\textbf{Light-weight}}  & ShuffleNet V2 & 0.021±0.031 & 0.015±0.029 & 0.017±0.033    
\end{tabular}
}
\end{table}

\begin{figure*}[t]
    \centering
    \includegraphics[width=0.95\linewidth]{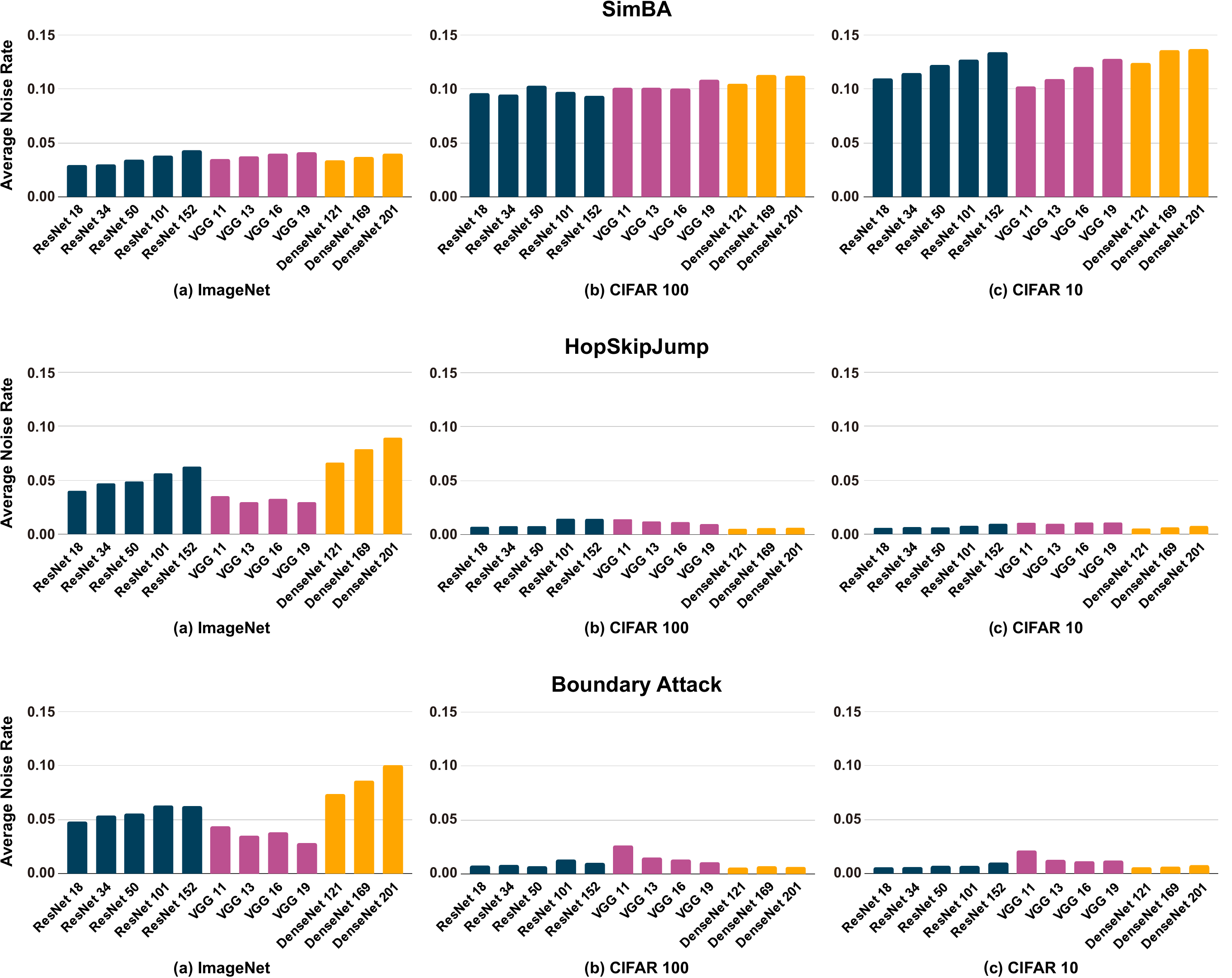}
   
    \caption{The impact of datasets on needed noise for SimBA, HopSkipJump, and Boundary attacks to succeed against 12 models from VGG, ResNet, and DensNet families. HopSkipJump and Boundary Attack seem to be more successful when the number of classes and input size are smaller. SimBA requires more noise against models trained on CIFAR-10 and CIFAR-100.}
    \label{fig:experiment3_1_diagram} 
   
\end{figure*} 

\subsection{{Exp 2}--- Model Diversity and Robustness} \label{sec:DiversityRobustness}
To explore the effects of architectural design on the robustness of models, we use various models, including MobileNet-V2, ShuffleNet-V2, Inception-V3, Xception, GoogLeNet, VGG-19, ResNet-152, and DenseNet-169. In this set of experiments, diversity refers to the underlying designs of the models and their associated impacts on the number of parameters. 

 MobileNet-V2 and ShuffleNet-V2 are designed to be lightweight models, suitable for deployment on mobile and embedded devices with limited computational resources. They use depth-wise separable convolutions to reduce the number of parameters. MobileNet-V2 has a higher accuracy than ShuffleNet-V2, but ShuffleNet-V2 is faster and has fewer parameters (see Tables \ref{tab:model_acc_2} and \ref{tab:model_info}). 
The Inception module was introduced by GoogLeNet (also known as Inception-V1), which allows for various filter sizes and reduces the number of parameters. It also contains auxiliary classifiers to aid in training. Based on the Inception architecture, Inception-V3, and Xception employ depth-wise separable convolutions to achieve high accuracy with fewer parameters.
The VGG-19 design is straightforward, consisting of stacked convolutional layers with small filter sizes followed by max-pooling layers. It has a large number of parameters and necessitates a substantial amount of computational power. ResNet-152 is a very deep model that employs residual connections to create very deep networks without vanishing gradients. The concept of dense connections, where each layer is linked to every other layer in a feed-forward manner, is the foundation of DenseNet-169. This results in feature reuse and a reduction in parameter count while increasing accuracy.
Table \ref{tab:model_info} shows the selected models to explore the effects of architectural designs on robustness. 

Depending on their parameter count, we group these models into three classes. \cib{1} \textit{Light-weight} models with an average of 2,891,738 parameters, including MobileNet-V2 and ShuffleNet-V2. \cib{2} \textit{Middle-weight} models such as Inception-V3, Xception, and GoogLeNet that have an average of 18,880,706 parameters. \cib{3} The rest of the models are \textit{Heavy-weight} with an average of 72,669,842
parameters.

Initially, and based on previous experiments (Exp 1 - Section \ref{sec:ComplexityRobustness}), the results show that the model complexity (\ie via the number of layers and parameters) would affect the robustness. Deeper networks tend to have higher resilience to perturbations in the input, whereas a larger number of parameters tends to make the model more susceptible to such perturbations. This set of experiments examines this assumption by employing a variety of models with varying numbers of parameters and designs to explore whether the design impacts robustness.

Figure \ref{fig:experiment2_diagram} shows that \textit{heavy-weight} models are actually less robust compared to their \textit{middle-weight} counterparts, as indicated by the average added noise. Considering only the number of parameters, the \textit{middle-weight} models are still more robust than the \textit{light-weight} despite the fewer parameters in \textit{light-weight} models. This suggests that the number of parameters is not necessarily indicative of robustness and that model designs matter in terms of robustness. \textit{Middle-weight} models, which feature various kernel sizes, auxiliary blocks, inception blocks, and stem designs, have been shown to be more robust. 
More robust models can be achieved through careful design choices rather than simply increasing the complexity of the model (more detailed results are in Table \ref{tab:experiment2_detail}).

\begin{observation*}{Model size and robustness}{}
Model architectures and the number of parameters do not guarantee robustness. The robustness of DNNs may vary across architectures, and the number of large parameters is not an indicator of robustness when multiple architectures are involved.  
\end{observation*}

\begin{figure*}[t]
    \centering
    \includegraphics[width=0.97\linewidth]{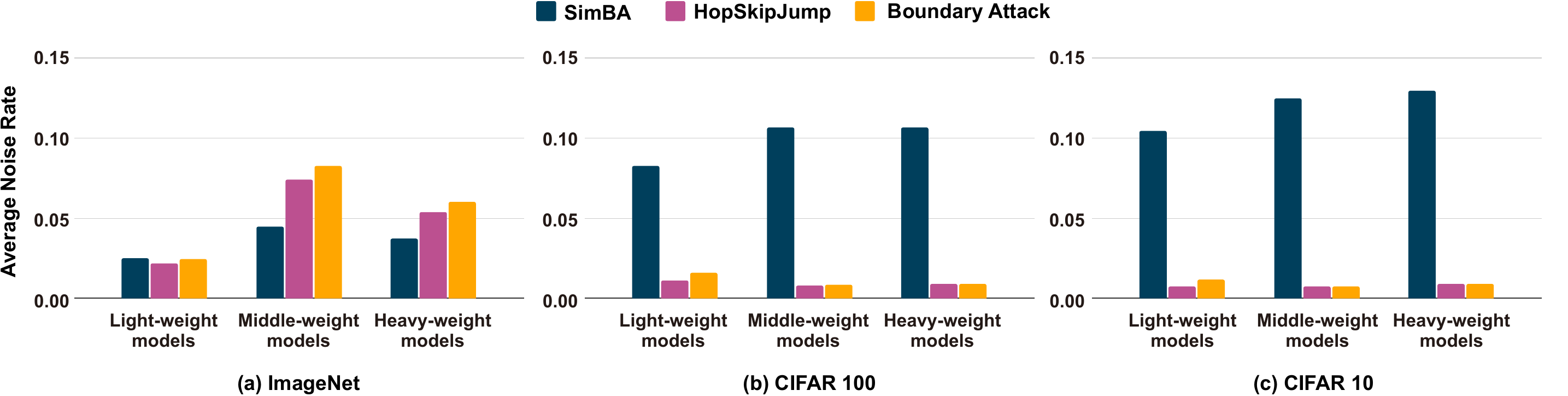}
   
    \caption{The impact of datasets on attacks targeting models with diverse architectural designs and sizes using different datasets.}
    \label{fig:experiment3_2_diagram} 
  
\end{figure*} 

\subsection{{Exp 3}--- Training Dataset and Robustness} 
In our earlier experiments, we explored the behavior of attacks on the ImageNet dataset. In this set of experiments, we explore the attack behavior with different models when using different datasets with varying input sizes and classes. To this end, we use ImageNet, CIFAR-100, and CIFAR-10, to examine the behavior of SimBA, HopSkipJump, and Boundary attacks against 12 models from the VGG, ResNet, and DenseNet families. Our results across different datasets are consistent with the results of previous experiments; \ie deeper networks within a model family tend to be more robust. Interestingly, the experiments with CIFAR-100 do not follow the same pattern. However, this might be due to the poor performance of models on CIFAR-100 as compared to their performance on other datasets (see Table \ref{tab:model_acc_1}), as adversarial misclassification can be partially caused by poor generalization.

\UlPara{Number of Classes and Robustness}
Based on our experiments, the success and stealthiness of adversarial attacks can depend on the specific attack strategy and the characteristics of the training dataset.
For example, HopSkipJump and Boundary Attack seem to be more successful when the number of classes is smaller. SimBA, on the other hand, requires more noise to succeed against all models. 
The primary idea behind a SimBA is to move the prediction from the benign class to the neighboring class. As a result, for the ImageNet dataset, SimBA perturbs considerably less noise than HopSkipJump as can be seen in Figure \ref{fig:experiment3_1_diagram}. The dense distribution of ImageNet's 1000 classes allows even a small amount of noise to change the prediction. However, due to the lower number of classes in the CIFAR-100 and CIFAR-10, SimBA introduces greater noise for input images to succeed.

\begin{observation*}{Number of classes and robustness}{}
  The small number of classes in a dataset has a significant negative influence on boundary-based attacks like SimBA. As the number of classes becomes less, the attack needs to add more noise. 
\end{observation*}

\UlPara{Input Size and Robustness}
Considering the results of HopSkipJump and Boundary Attack, the needed perturbations to succeed are significantly lower when attacking models trained in the CIFAR-10 and CIFAR-100 datasets than when attacking models trained on ImageNet (see Figures \ref{fig:experiment3_1_diagram} and \ref{fig:experiment3_2_diagram}). The results suggest that both the number of classes and the size of the input may influence the robustness of the models. Additionally, evaluating the robustness of models across different datasets and input sizes is crucial to ensuring their reliability in real-world applications.

Generally, when using models with diverse architectural designs and sizes, the behavior of the attack depends on the dataset and attack strategy. For example, Figure \ref{fig:experiment3_2_diagram} shows that \textit{light-weight} models need more perturbations than \textit{middle-weight} and \textit{heavy-weight} models when adopting HopSkipJump and Boundary Attack on CIFAR datasets. On the other hand, using SimBA, the attack behavior is consistent with the patterns observed in ImageNet (\ie model design is crucial to robustness as \textit{middle-weight} models tend to be more resilient than others to adversarial perturbations).

   \begin{observation*}{Size of input and robustness}{}
  The input image size has a significant impact on the performance of black-box attacks (HopSkipJump and Boundary attacks). Smaller images are generally more vulnerable to adversarial attacks due to their lower resolution and simpler features.
\end{observation*}

\subsection{{Exp 4}--- Defenses against Black-box Attacks}  
Arguably, defending against an attempted attack is as crucial as predicting an accurate result. The purpose of defense techniques is to minimize the effects of adversarial perturbations and protect models against adversarial attacks. 

\begin{figure}[t]
    \centering
    \includegraphics[width=\linewidth]{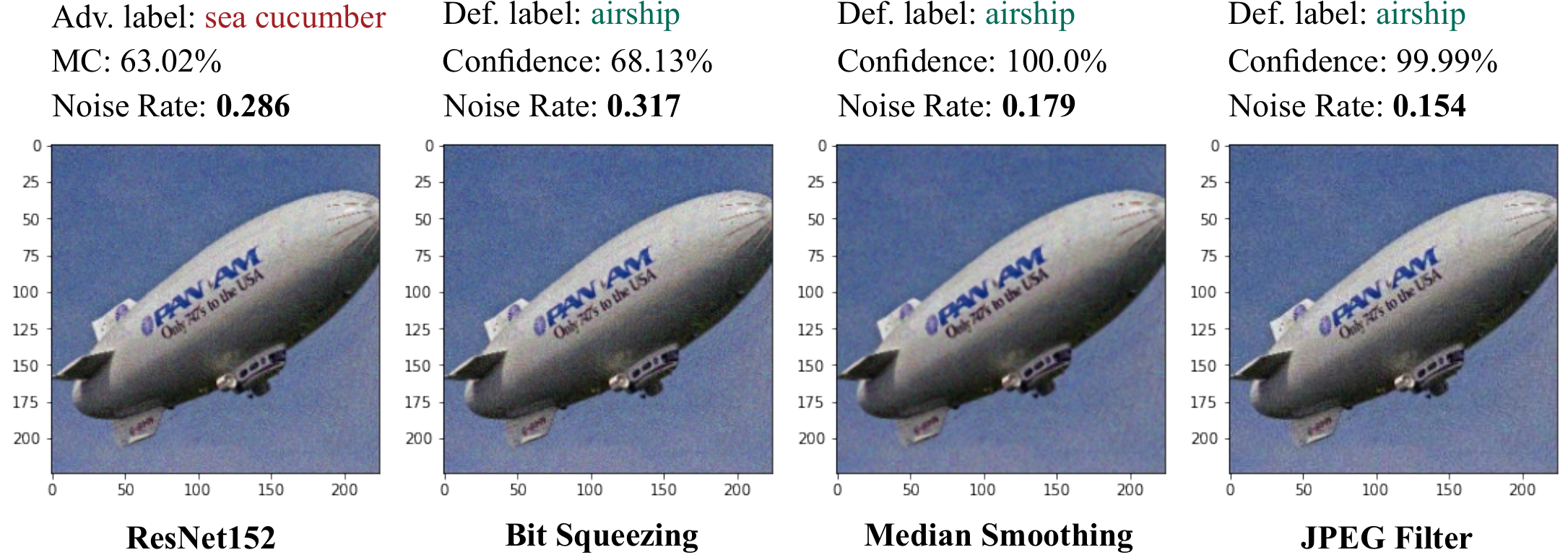}
    \caption{Bit squeezing, median, and JPEG defensive techniques and the needed amount of noise to defend the model effectively. Under the HopSkipJump attack, bit squeezing seems to add more noise than median and JPEG filters, reducing its confidence score.}
    \label{fig:defense_example} 
   
\end{figure}

\UlPara{Preprocessor-type Defenses} In this set of experiments, we explore the effects of applying preprocessor-type defenses against black-box attacks. Using Bit Squeezing, Median Smoothing, and the JPEG Filter, our results show that most attacks are ineffective even with minimal parameter settings for defenders. 
The reason for this is that most adversarial attacks, such as boundary and SimBA attacks, stop perturbing noise after a class crosses its boundary and moves to another class region. Therefore, such attacks result in a low MC score, and adversarial images revert to the original class border with minimal defender parameters. For example, Figure \ref{fig:defense_example} shows that ResNet-152 misclassifies the airship as a sea cucumber with an MC score of 63.02\% after adding 28.6\% noise under the HopSkipJump attack. 
When using the bit-squeezing defensive technique (with 31.7\% corrective noise), the model correctly classifies the image with a confidence score of 68.13\%. 
The median smoothing and JPEG filtering techniques allow the model to correctly classify the image with a confidence score of $\approx 100\%$. 

\begin{figure*}[t]
    \centering
    \includegraphics[width=0.9\linewidth]{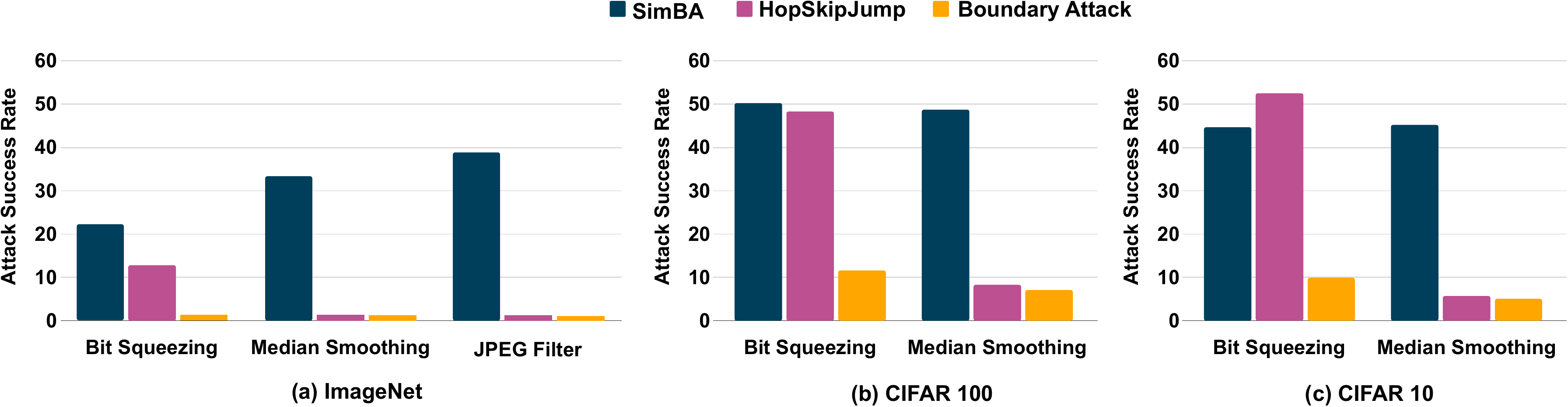}
   
    \caption{The effectiveness of the various defensive strategies and observes how they bring the percentage of successful attacks down from 100\%. Note that JPEG defense could not be applicable to CIFAR dataset due to the smaller size of the images.}
    \label{fig:experiment_4_defenses} 
   
\end{figure*}  

Figure \ref{fig:experiment_4_defenses} depicts the performance of each defense technique in terms of decreasing the attack success rate (from the initial 100\%). The results show that\cib{1} Boundary attacks are ineffective across all defenses and datasets; \cib{2} Against SimBA, no defense has shown remarkable reductions regardless of the dataset; \cib{3} Other defenses are more efficient than bit-squeezing against HopSkipJump attack.

    \begin{observation*}{Weakness of black-box attacks}{}
  Even with weak defensive parameters, black-box attacks are vulnerable to countermeasures.  
\end{observation*}

\begin{figure}[t]
    \centering
    \includegraphics[width=0.6\linewidth]{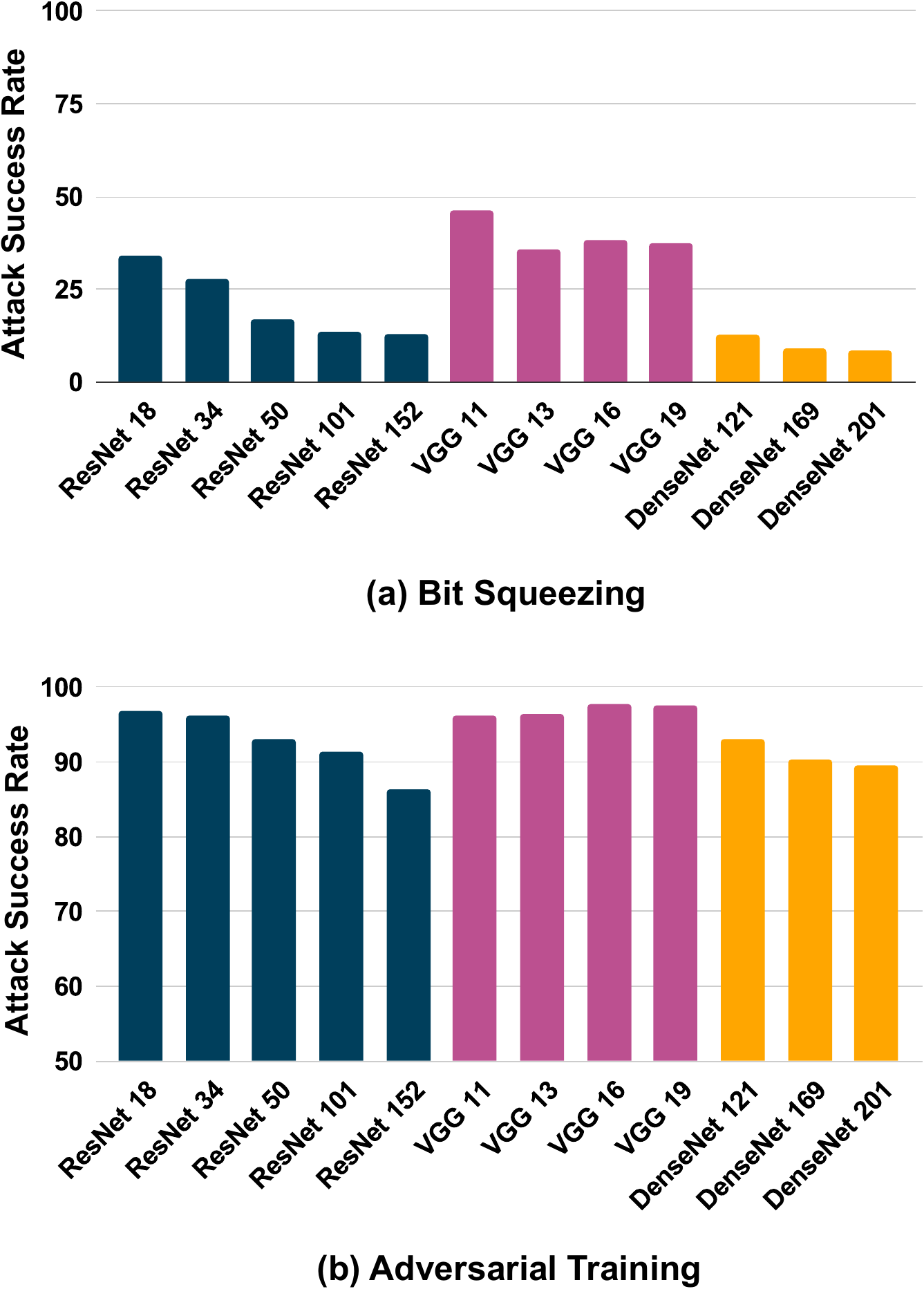}
   
    \caption{Bit Squeezing defense and Adversarial Training against SimBA on ImageNet. The model becomes easier to defend as the number of layers on the same model increases}
    \label{fig:trend_bit_adversarial_training_results} 
   
\end{figure}

\UlPara{Adversarial Training Defenses} 
Figure \ref{fig:trend_bit_adversarial_training_results} (a) shows an interesting observation, which is that as the number of layers in a model increases, the defense of the model becomes easier with preprocessor-type defenses. 
Furthermore, in addition to evaluating the models on the standard training data, we also conducted adversarial training. This process includes training the models with 500 adversarial samples. The results show that adversarial training (as shown in Figure \ref{fig:trend_bit_adversarial_training_results} (b)) helps the model against attacks.

    \begin{observation*}{Number of layers affects robustness}{}
  The model becomes easier to defend as the number of layers on the same model increases (See Figure \ref{fig:trend_bit_adversarial_training_results}).
\end{observation*}

\section{Discussion and Open Challenges} \label{sec:discussion}
\BfPara{Key Takeaways} With our experimental study, we have reported several insights and recommendations for developing new attack proposals based on our comprehensive analysis. As indicated by the results of Experiment 1, the complexity of a model is directly proportional to the amount of additive noise required for the generation of adversarial examples. 
For an effective attack on DNN models on black-box settings, the attack mechanism should keep the noise visually undetectable and circumvents defensive mechanisms. Moreover, the attack must take into account not just being query efficient and having a high attack success rate, but also consider an effective way of adding noise, controlling the amount of noise, and selecting the proper regions for adding noise.

Experiment 2 demonstrates that enhancing the model's robustness is achievable by closely examining the model's internal structure. We found that black-box attacks, \ie HopSkipJump, and boundary attacks are highly dependent on the size of the input images, while SimBA is sensitive to the number of classes in the dataset. As for defenses, preprocessor-based defenses hinder the effectiveness of boundary attacks, while adversarial training significantly contributes to enhancing the model's robustness.

\vspace{1.5ex}
\BfPara{Comparison} 
Images are the most commonly used for evaluating attacks and defenses. Initially, the MNIST and CIFAR datasets are used to assess the vulnerability of DNN to adversarial instances in the image domain \cite{krizhevsky2009learning}. In recent years, the majority of research efforts have favored ImageNet over MNIST and CIFAR. Using the ImageNet hierarchical class system, Ozbulak \etal \cite{ozbulak2021evaluating} conducted a comprehensive investigation of the characteristics of the classes into which adversarial instances are misclassified. Specifically, it focuses on model-to-model adversarial transferability and misclassification classes using two of the most often used adversarial approaches, five distinct DNN architectures (AlexNet, SqueezeNet, VGG-16, ResNet, and DenseNet), and two vision transformer architectures (ViT-Base and ViT-Large). Gragnaniello \etal \cite{gragnaniello2018analysis} examine the effectiveness of: \cib{1} adversarial attacks, including gradient descent, adversarial noise, and GAN-based attacks, and \cib{2} CNN-based detectors such as SPAM+SVM, Bayar2016, Cozzolino2017, and very deep nets.  

Unlike prior studies, which have primarily focused on creating more effective adversary perturbation or assessing adversarial countermeasures, we address a previously underexplored problem: \textit{analyzing black-box attacks across different datasets using a range of DL models}. Five evaluation matrices, three state-of-the-art datasets, and a wide range of DNN models are included in the initial stage of the current research. Obviously, the preliminary findings are merely proof of concept, and additional in-depth investigations are required to have a complete grasp of the issues. 

\section{Conclusion} \label{sec:conclusion}
This work presents the results of large-scale experiments on the effect of adversarial attacks and countermeasures on distinct models utilizing three state-of-the-art datasets. The results of the experiments revealed that there is a relationship between model complexity and robustness and that having a large number of parameters does not imply that the model is more robust. 
Black-box attacks are susceptible to the kind of dataset used in terms of input size and the number of classes. For HopSkipJump and boundary black-box attacks, the defensive technique with minimal parameter settings is effective. Compared to the median and JPEG filters, the bit-squeezing defensive technique generated more noise, lowering the confidence score.
Future directions can explore the behavior of the white-box and gray-box attacks, including recent advanced attacks such as \cite{abdukhamidov2021advedge, abdukhamidov2023hardening, zhang2020interpretable} with other defense strategies. Moreover, studying the complexity and robustness of various architectures beyond CNNs, such as Vision Transformers, is an interesting future direction.

\section*{Acknowledgment}
This work was supported by the National Research Foundation of Korea(NRF) grant funded by the Korea government(MSIT)(No. 2021R1A2C1011198), (Institute for Information \& communications Technology Planning \& Evaluation) (IITP) grant funded by the Korea government (MSIT) under the ICT Creative Consilience Program (IITP-2021-2020-0-01821), and AI Platform to Fully Adapt and Reflect Privacy-Policy Changes (No. 2022-0-00688).

\balance
\bibliographystyle{plain}
\bibliography{main}

\begin{thebibliography}{10}

\bibitem{abdukhamidov2021advedge}
Eldor Abdukhamidov, Mohammed Abuhamad, Firuz Juraev, Eric Chan-Tin, and Tamer AbuHmed.
\newblock Advedge: Optimizing adversarial perturbations against interpretable deep learning.
\newblock In {\em International Conference on Computational Data and Social Networks}, pages 93--105. Springer, 2021.

\bibitem{abdukhamidov2023hardening}
Eldor Abdukhamidov, Mohammed Abuhamad, Simon~S Woo, Eric Chan-Tin, and Tamer Abuhmed.
\newblock Hardening interpretable deep learning systems: Investigating adversarial threats and defenses.
\newblock {\em IEEE Transactions on Dependable and Secure Computing}, 2023.

\bibitem{abdukhamidov2023unveiling}
Eldor Abdukhamidov, Mohammed Abuhamad, Simon~S Woo, Eric Chan-Tin, and Tamer Abuhmed.
\newblock Unveiling vulnerabilities in interpretable deep learning systems with query-efficient black-box attacks.
\newblock {\em arXiv preprint arXiv:2307.11906}, 2023.

\bibitem{abuhamad2021large}
Mohammed Abuhamad, Tamer Abuhmed, David Mohaisen, and Daehun Nyang.
\newblock Large-scale and robust code authorship identification with deep feature learning.
\newblock {\em ACM Transactions on Privacy and Security (TOPS)}, 24(4):1--35, 2021.

\bibitem{ali2022effective}
Sajid Ali, Omar Abusabha, Farman Ali, Muhammad Imran, and Tamer Abuhmed.
\newblock Effective multitask deep learning for iot malware detection and identification using behavioral traffic analysis.
\newblock {\em IEEE Transactions on Network and Service Management}, 2022.

\bibitem{brendel2018decision}
Wieland Brendel, Jonas Rauber, and Matthias Bethge.
\newblock Decision-based adversarial attacks: Reliable attacks against black-box machine learning models.
\newblock In {\em 6th International Conference on Learning Representations, {ICLR} 2018, April 30 - May 3, 2018, Conference Track Proceedings}, pages 1--12, 2018.

\bibitem{carlini2016hidden}
Nicholas Carlini, Pratyush Mishra, Tavish Vaidya, Yuankai Zhang, Micah Sherr, Clay Shields, David Wagner, and Wenchao Zhou.
\newblock Hidden voice commands.
\newblock In {\em 25th USENIX Security Symposium}, 2016.

\bibitem{carlini2017adversarial}
Nicholas Carlini and David Wagner.
\newblock Adversarial examples are not easily detected: Bypassing ten detection methods.
\newblock In {\em Proceedings of the 10th ACM workshop on artificial intelligence and security}, pages 3--14, 2017.

\bibitem{chen2020hopskipjumpattack}
Jianbo Chen, Michael~I Jordan, and Martin~J Wainwright.
\newblock Hopskipjumpattack: A query-efficient decision-based attack.
\newblock In {\em 2020 ieee symposium on security and privacy (sp)}, pages 1277--1294, San Francisco, CA, USA, 2020. IEEE.

\bibitem{dziugaite2016study}
Gintare~Karolina Dziugaite, Zoubin Ghahramani, and Daniel~M Roy.
\newblock A study of the effect of jpg compression on adversarial images.
\newblock {\em arXiv preprint arXiv:1608.00853}, pages 1--8, 2016.

\bibitem{gragnaniello2018analysis}
Diego Gragnaniello, Francesco Marra, Giovanni Poggi, and Luisa Verdoliva.
\newblock Analysis of adversarial attacks against cnn-based image forgery detectors.
\newblock In {\em 2018 26th European Signal Processing Conference (EUSIPCO)}, pages 967--971. IEEE, 2018.

\bibitem{guo2019simple}
Chuan Guo, Jacob Gardner, Yurong You, Andrew~Gordon Wilson, and Kilian Weinberger.
\newblock Simple black-box adversarial attacks.
\newblock In {\em International Conference on Machine Learning}, pages 2484--2493, Honolulu, HI, USA, 2019. PMLR.

\bibitem{he2016deep}
Kaiming He, Xiangyu Zhang, Shaoqing Ren, and Jian Sun.
\newblock Deep residual learning for image recognition.
\newblock In {\em Proceedings of the IEEE conference on computer vision and pattern recognition}, pages 770--778, Las Vegas Nevada, 2016. IEEE Xplore.

\bibitem{10397075}
Wei Jia, Zhaojun Lu, Runze Yu, Liaoyuan Li, Haichun Zhang, Zhenglin Liu, and Gang Qu.
\newblock Fooling decision-based black-box automotive vision perception systems in physical world.
\newblock {\em IEEE Transactions on Intelligent Transportation Systems}, pages 1--12, 2024.

\bibitem{krizhevsky2009learning}
Alex Krizhevsky, Geoffrey Hinton, et~al.
\newblock Learning multiple layers of features from tiny images.
\newblock {\em Technical report}, pages 1--60, 2009.

\bibitem{kurakin2016adversarial}
Alexey Kurakin, Ian Goodfellow, and Samy Bengio.
\newblock Adversarial machine learning at scale.
\newblock {\em arXiv preprint arXiv:1611.01236}, pages 1--17, 2016.

\bibitem{lecun1998gradient}
Yann LeCun, L{\'e}on Bottou, Yoshua Bengio, and Patrick Haffner.
\newblock Gradient-based learning applied to document recognition.
\newblock {\em Proceedings of the IEEE}, 86(11):2278--2324, 1998.

\bibitem{osadchy2017no}
Margarita Osadchy, Julio Hernandez-Castro, Stuart Gibson, Orr Dunkelman, and Daniel P{\'e}rez-Cabo.
\newblock No bot expects the deepcaptcha! introducing immutable adversarial examples, with applications to captcha generation.
\newblock {\em IEEE Transactions on Information Forensics and Security}, 12(11):2640--2653, 2017.

\bibitem{ozbulak2021evaluating}
Utku Ozbulak, Maura Pintor, Arnout Van~Messem, and Wesley De~Neve.
\newblock Evaluating adversarial attacks on imagenet: A reality check on misclassification classes.
\newblock {\em arXiv preprint arXiv:2111.11056}, pages 1--16, 2021.

\bibitem{russakovsky2015imagenet}
Olga Russakovsky, Jia Deng, Hao Su, Jonathan Krause, Sanjeev Satheesh, Sean Ma, Zhiheng Huang, Andrej Karpathy, Aditya Khosla, Michael Bernstein, et~al.
\newblock Imagenet large scale visual recognition challenge.
\newblock {\em International journal of computer vision}, 115(3):211--252, 2015.

\bibitem{shafahi2019adversarial}
Ali Shafahi, Mahyar Najibi, Mohammad~Amin Ghiasi, Zheng Xu, John Dickerson, Christoph Studer, Larry~S Davis, Gavin Taylor, and Tom Goldstein.
\newblock Adversarial training for free!
\newblock {\em Advances in Neural Information Processing Systems}, 32, 2019.

\bibitem{simonyan2014very}
Karen Simonyan and Andrew Zisserman.
\newblock Very deep convolutional networks for large-scale image recognition.
\newblock {\em arXiv preprint arXiv:1409.1556}, pages 1--14, 2014.

\bibitem{singh2021classification}
Jaiteg Singh, Deepak Thakur, Tanya Gera, Babar Shah, Tamer Abuhmed, and Farman Ali.
\newblock Classification and analysis of android malware images using feature fusion technique.
\newblock {\em IEEE Access}, 9:90102--90117, 2021.

\bibitem{szegedy2014intriguing}
Christian Szegedy, Wojciech Zaremba, Ilya Sutskever, Joan Bruna, Dumitru Erhan, Ian Goodfellow, and Rob Fergus.
\newblock Intriguing properties of neural networks.
\newblock In {\em International Conference on Learning Representations, {ICLR} 2014}, pages 1--10, 2014.

\bibitem{wang2020mgaattack}
Lina Wang, Kang Yang, Wenqi Wang, Run Wang, and Aoshuang Ye.
\newblock Mgaattack: Toward more query-efficient black-box attack by microbial genetic algorithm.
\newblock In {\em Proceedings of the 28th ACM International Conference on Multimedia}, pages 2229--2236, 2020.

\bibitem{wang2004ssim}
Zhou Wang, A.C. Bovik, H.R. Sheikh, and E.P. Simoncelli.
\newblock Image quality assessment: from error visibility to structural similarity.
\newblock {\em IEEE Transactions on Image Processing}, 13(4):600--612, 2004.

\bibitem{xu2018feature}
Weilin Xu, David Evans, and Yanjun Qi.
\newblock Feature squeezing: Detecting adversarial examples in deep neural networks.
\newblock In {\em 25th Annual Network and Distributed System Security Symposium, {NDSS} 2018, February 18-21, 2018}, pages 1--15, HSan Diego, California, USA, 2018. The Internet Society.

\bibitem{zhang2017dolphinattack}
Guoming Zhang, Chen Yan, Xiaoyu Ji, Tianchen Zhang, Taimin Zhang, and Wenyuan Xu.
\newblock Dolphinattack: Inaudible voice commands.
\newblock In {\em Proceedings of the 2017 ACM SIGSAC Conference on Computer and Communications Security}, pages 103--117, 2017.

\bibitem{zhang2020interpretable}
Xinyang Zhang, Ningfei Wang, Hua Shen, Shouling Ji, Xiapu Luo, and Ting Wang.
\newblock Interpretable deep learning under fire.
\newblock In {\em 29th USENIX Security Symposium}, 2020.

\end{thebibliography}


\end{document}